\documentclass[%
 reprint,
superscriptaddress,
showpacs,
 amsmath,amssymb,
 aps,prl,
 citeautoscript
]{revtex4-1}

\usepackage{graphicx}% Include figure files
\graphicspath{{pict/}{}}

\usepackage{hyperref}
\hypersetup{pdfstartview={FitH},pdfpagemode={UseNone},
            colorlinks,linkcolor=blue, citecolor=blue, urlcolor=blue,
            bookmarksopen=true, pdfnewwindow=true}

\usepackage{dcolumn}% Align table columns on decimal point
\usepackage{bm}% bold math

\begin{document}

\title{Observation of localized modes at effective gauge field interface in synthetic photonic lattice}% Force line breaks with \\

\author{Artem~V. Pankov}
\affiliation{Novosibirsk State University, Pirogova str. 2, Novosibirsk 630090, Russia}

\author{Ilya~D. Vatnik}%
\affiliation{Novosibirsk State University, Pirogova str. 2, Novosibirsk 630090, Russia}

\author{Dmitry~V. Churkin}%
\affiliation{Novosibirsk State University, Pirogova str. 2, Novosibirsk 630090, Russia}

\author{Andrey~A. Sukhorukov}
\email{Andrey.Sukhorukov@anu.edu.au}
\affiliation{Nonlinear Physics Centre, Research School of Physics and Engineering, Australian National University, Canberra, ACT 2601, Australia}

\date{\today}% It is always \today, today,

\begin{abstract}
We predict a generic mechanism of wave localization at an interface between uniform gauge fields, arising due to propagation-dependent phase accumulation similar to Aharonov-Bohm phenomenon. We realize experimentally a photonic mesh lattice with real-time control over the vector gauge field, and observe robust localization under a broad variation of gauge strength and direction, as well as structural lattice parameters. This suggests new possibilities for confining and guiding waves in diverse physical systems through the synthetic gauge fields.
\end{abstract}

%\pacs{42.81.Qb, 42.25.Gy, 03.65.Vf}
\maketitle

The recent proposals and experimental demonstrations of artificial gauge potential open new opportunities for the manipulation of neutral particles such as photons~\cite{Fang:2012-782:NPHOT, Tzuang:2014-701:NPHOT}. The effective magnetic field arises when the waves accumulate a phase which depends on the propagation direction, representing Aharonov-Bohm effect~\cite{Fang:2012-153901:PRL, Longhi:2014-5892:OL, Li:2014-3225:NCOM}, which can be implemented for photons through the specially introduced dynamic modulation in lattices of coupled waveguides or resonators. These concepts underpin the realization of a broad variety of fundamental phenomena including dynamic localization~\cite{Garanovich:2012-1:PRP, Yuan:2015-243901:PRL}, robust scattering-immune one-way edge states and topological insulators~\cite{Rechtsman:2013-196:NAT, Sounas:2017-774:NPHOT} in non-magnetic materials.

The presence of a gauge field can fundamentally modify the wave localization even in topologically trivial cases~\cite{Garanovich:2012-1:PRP}. It was recently suggested that one-way modes in nonreciprocal waveguides can be realized by introducing a uniform gauge field in the waveguide core~\cite{Lin:2014-31031:PRX}. A spatially uniform and time-independent gauge does not induce a magnetic field and topology in the core remains unchanged. Nevertheless, it can lead to a direction-dependent shift of the optical dispersion~\cite{Fang:2012-782:NPHOT} such that there appears an effective refractive index contrast between the core and cladding, which can support modes according to the usual waveguiding concept.

In this work, we predict and observe experimentally a new regime of mode localization at a boundary between two regions with different uniform gauge fields. Such configuration induces a magnetic field only at the interface, and we find that it can support strongly confined optical modes. Remarkably, localization occurs even when the refractive index contrast is zero, breaking away from the usual conventions of index-guiding and thus allowing for an extra degree of flexibility in wave control.
This general principle can be applied to various optical structures, as well as other physical systems with engineered gauge fields including cold atoms~\cite{Gross:2017-995:SCI}, and exciton-polaritons~\cite{Lim:2017-14540:NCOM}.

\begin{figure}[htbp]
\centering
\includegraphics[width=\columnwidth]{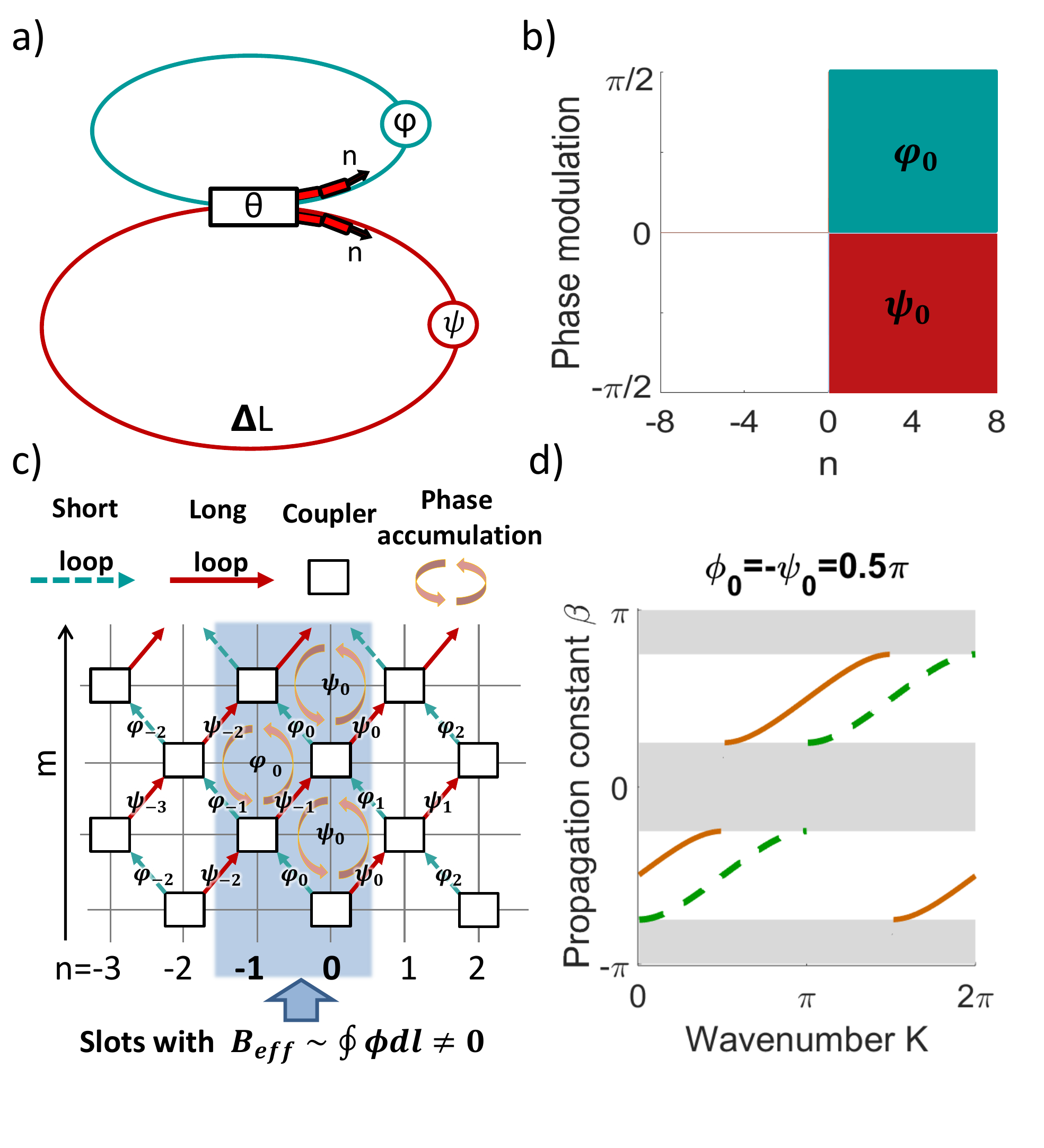}
\caption{
a)~Synthetic mesh lattice realized with two coupled fiber loops.
b)~Phase shifts $\phi(n)$ in short and $\psi(n)$ in long loops applied to create a gauge field interface.
c)~Scheme of phase accumulation on a mesh lattice graph and appearance of an effective magnetic field in the shaded area at the interface.
d)~Mode dispersion $\beta(K)$ on different sides of the interface for
$\theta=0.25\pi$. Green dash line~-- conventional case with no phase modulation and zero gauge field for $n<0$. Orange line~-- out-of-phase modulation profile as shown on panel (b), inducing a transverse gauge field and wavenumber shift $\Delta K$. Shading indicates band-gaps.
}
\label{fig:MeshLatticeScheme}
\end{figure}

We demonstrate the localization in a synthetic mesh lattice (SML).
Such types of lattices were employed to demonstrate a number of fundamental effects including discrete quantum walks~\cite{Schreiber:2011-180403:PRL}, Bloch oscillations \cite{Schreiber:2010-50502:PRL}, Anderson localization \cite{Schreiber:2011-180403:PRL, Vatnik:2017-4301:SRP}, Parity-Time (PT) symmetry breaking \cite{Regensburger:2012-167:NAT, Miri:2012-23807:PRA}, discrete solitons \cite{Wimmer:2015-7782:NCOM}, optical diametric drive acceleration through action-reaction symmetry breaking \cite{Wimmer:2013-780:NPHYS}, and defect states in PT-symmetric environment \cite{Regensburger:2013-223902:PRL}. In this work, we suggest and implement experimentally an essential extension of previous schemes to enable fully flexible realization of vector gauge fields with arbitrary magnitude and direction, enabling the observation of new localization regimes.

We employ an optical-fiber based system comprising two loops connected via a directional coupler, which strength is characterized by parameter $\theta$, see Fig.~\ref{fig:MeshLatticeScheme}(a). The two fiber loops differ in length by $\Delta L$, and a single pulse launched into the system produces a train of pulses circulating, interfering and multiplying within the loops, which is theoretically equivalent to the dynamics in a mesh lattice~\cite{Regensburger:2012-167:NAT}.
The overall evolution depends on the loss or amplification and phase shifts acquired by each single pulse during the roundtrip in the respective fiber loop. The optical phases can be precisely controlled by means of electro-optical modulators together with electrical waveform generators. Whereas in previous studies the phase shifts were only modulated in one fiber loop, we reveal that the phase modulation in both loops is required for the full control over the synthetic vector gauge field.

We generalize the theory of light dynamics in temporal mesh lattices~\cite{Regensburger:2012-167:NAT} by considering the independent control of phases in both loops, and formulate equations for the amplitudes in the short ($u_n^m $) and long ($v_n^m $) loops immediately before the coupler:
\begin{equation} \label{model}
\begin{aligned}
u^{m+1}_{n} = \exp( i \phi_{n} ) \left[\cos(\theta) u^{m}_{n+1} + i\sin(\theta) v^{m}_{n+1} \right] ,
\\
v^{m+1}_{n} = \exp( i \psi_{n} ) \left[ \cos(\theta)v^{m}_{n-1} + i\sin(\theta) u^{m}_{n-1} \right] .
\end{aligned}
\end{equation}
Here $\theta$ defines the coupling ratio of the fiber splitter, $m$ is the roundtrip number (discrete time coordinate), and $n$ is a space-like coordinate defined by particular position of a pulse within the loop. The phase modulation is described by $\phi_n$ in the short and $\psi_n$ in the long fiber loops. We assume that losses are compensated by amplifiers, such that their presence can be neglected under practical experimental conditions discussed in the following.

We first consider the effect of constant phase modulation $(\phi_n=\phi_0, \psi_n=\psi_0)$, and show that this regime corresponds to a uniform gauge field.
We seek solutions for the eigenmodes in the form of Bloch-like functions,
\begin{equation} \label{eq:BlochLikeEigenmodes}
   u_n^m=U\exp(iKn+im\beta), \, v_n^m=V\exp(iKn+im\beta),
\end{equation}
where $\beta$ is the propagation constant proportional to a longitudinal wavenumber, and $K$ is the transverse wavenumber.
We find that the mode spectrum
consists of two bands,
\begin{equation} \label{eq:betageneral}
    \beta_\pm=\frac{\psi_0+\phi_0}{2}\pm \cos^{-1}[\cos(K-\psi_0/2+\phi_0/2)\cos(\theta)] .
\end{equation}
We see that the phase modulation has a twofold effect. On the one hand, it leads to a shift of propagation constant by $\Delta \beta = (\psi_0+\phi_0)/ 2$, similar to a modification of refractive index. On the other hand, there can appear a transverse wavenumber shift $\Delta K = (\psi_0-\phi_0)/ 2$. Based on the general theory of dispersion in synthetic lattices~\cite{Lin:2014-31031:PRX}, we conclude that the constant phase modulations introduce a uniform synthetic gauge field
\begin{equation} \label{eq:gauge}
    {\bf A} = (\Delta K, \Delta \beta)
            = \left( \frac{\psi_0-\phi_0}{2}, \frac{\psi_0+\phi_0}{2} \right).
\end{equation}
This demonstrates the key importance of modulating the phases in both fiber loops to control the magnitude and direction of the gauge field. In particular, for out-of-phase modulation with $\psi_0 = -\phi_0$ there appears effective transverse gauge field ($\Delta K \ne 0$) yet no band-gap shift ($\Delta \beta = 0$). In contrast, in previously studied synthetic photonic lattices with phase modulation in one loop only (i.e. $\phi_n \equiv 0$), the direction of the gauge field was rigidly fixed to one direction with $\Delta K \equiv \Delta \beta$.

\begin{figure}[htbp]
\centering
{\includegraphics[width=\columnwidth]{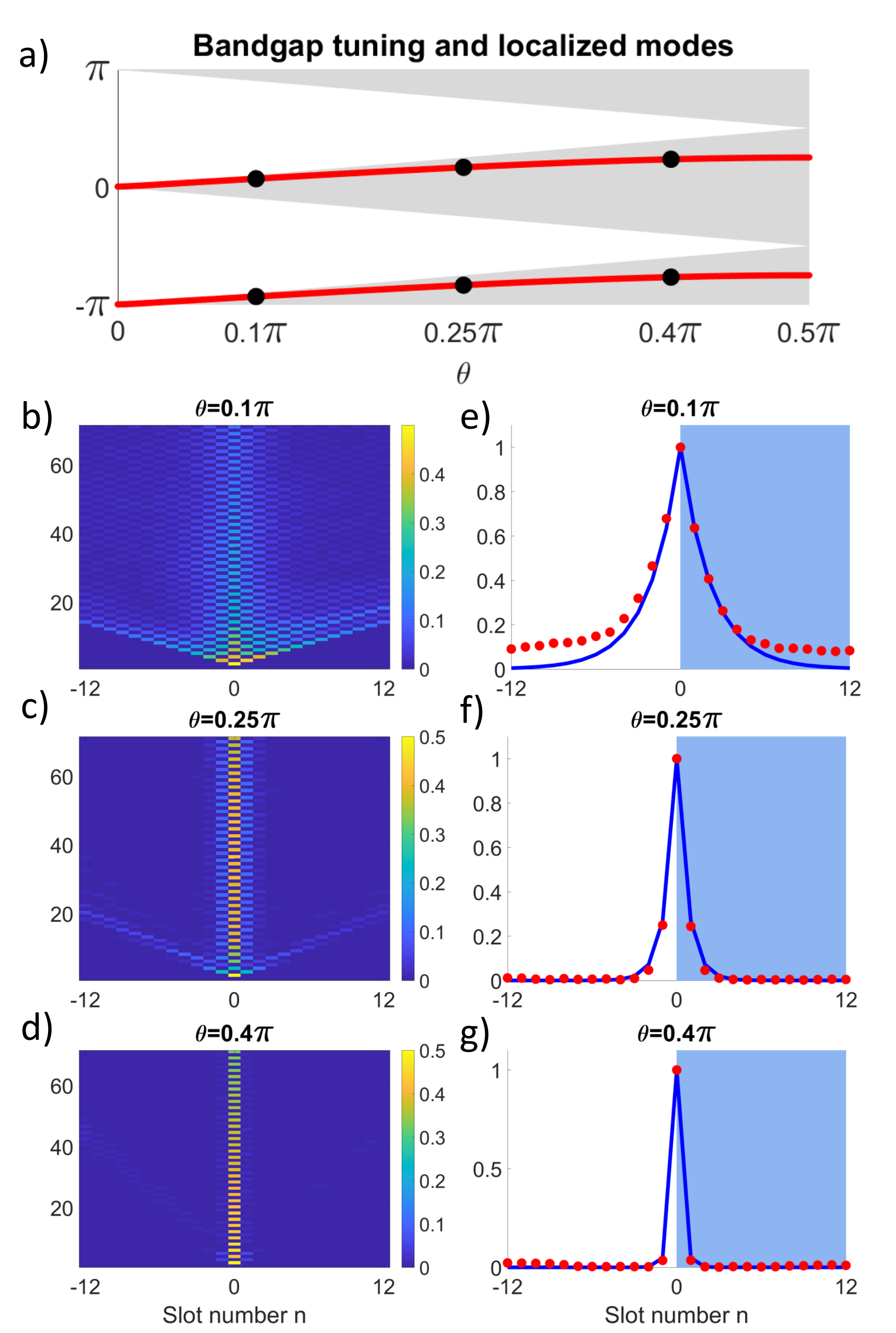}}
\caption{
(a)~Bandgaps (shaded regions) and propagation constants $\beta$ for localized modes (red lines) vs. the loop coupling $\theta$. Black dots correspond to experimental conditions in the following plots.
(b-d)~Experimentally measured evolution of the light pulses in the long loop for different $\theta$ as indicated by labels.
(e-g)~Red dots~-- experimentally found intensity distributions for modes localized at interface extracted from the corresponding images on the left. Blue lines~-- theoretical predictions.
For all the plots, $\phi_0=-\psi_0=0.5\pi$.
}
\label{fig:fig2}
\end{figure}

Next, we consider the effect of a gauge field jump, which creates an effective magnetic field just at the interface.
With no loss of generality, we analyze the phase modulation profile with $\phi_n = \psi_n = 0$ at $n < 0$, while $\phi_n = \phi_0$ and $\psi_n=\psi_0$ at $n \ge 0$, see Fig.~\ref{fig:MeshLatticeScheme}(b). We illustrate the corresponding phase accumulation on a mesh lattice graph in Fig. \ref{fig:MeshLatticeScheme}(c).
The effective magnetic field $B_{\rm eff}$ is proportional to the total phase accumulation around the closed paths enclosing the sets of four nearest beam-splitters. It vanishes for $n\leq-2$ and $n\geq 1$ as in these regions the phase modulation and the associated gauge fields are constant.
In contrast, an effective magnetic field appears next to the gauge field interface, see the shaded area in Fig. \ref{fig:MeshLatticeScheme}(c). In this region, the phase accumulation depends on the order of pulse propagation through the loops, i.e. it is different for short-long or long-short paths,
analogous to the Aharonov-Bohm effect.

We find that the gauge field discontinuity supports localized modes facilitated by the effective magnetic fields at the interface.
This is most clearly evident when the bandgaps are identical on both sides of the interface, for $\phi_0 = -\psi_0$, as illustrated in Fig.~\ref{fig:MeshLatticeScheme}(d).
There exist two localized modes with propagation constants placed inside the bandgaps for any coupling coefficient $\theta$, see Fig.~\ref{fig:fig2}(a). These modes are related by a transformation which is an invariant of the model Eqs.~(\ref{model}),
\begin{equation} \label{modes2}
  \{u,v\}_n \exp[i \beta m] \rightarrow  \{u,v\}_n (-1)^n \exp[i (\beta+\pi) m] .
\end{equation}
We see that the modes have the same intensity profiles, but different phases and propagation constants offset by $\pi$.
The modes have exponentially localized tails, see characteristic theoretical profiles shown with solid lines in Figs.~\ref{fig:fig2}(e-g). Remarkably, these modes are supported purely by effective magnetic field at the interface. Previously, effective magnetic field was reported to produce waveguiding effect due to bandgap shifts yet no interface states have been identified~\cite{Lin:2014-31031:PRX}.

We experimentally realize synthetic mesh lattice using two coupled fiber loops with losses compensated by amplifiers~\cite{Regensburger:2011-233902:PRL, Schreiber:2010-50502:PRL, Miri:2012-23807:PRA, Vatnik:2017-4301:SRP}. The initial pulse launched into the short loop multiplies at the coupler each roundtrip, resulting in a nontrivial time dependence of the intensity measured by a photodetector in the long loop. %Cutting this time trace on pieces corresponding to different roundtrips and putting pulses appearing
We measure the pulses with a photodiode, and visualize the
evolution on a mesh lattice. To realize the gauge fields, electro-optical modulators are placed into each loop. The modulators are programmed to produce the type of phase modulation patterns illustrated in Fig.~\ref{fig:MeshLatticeScheme}(b).
Details about the setup are provided in Sec.~1 of Supplemental Material~\cite{SupplementalMaterial}.
To study the light dynamics at the gauge field jump, we launch a single pulse in the short loop at the interface position $n=0$ (other excitation positions are analyzed in Sec.~2 of Supplemental Material~\cite{SupplementalMaterial}).
This results in a simultaneous excitation of both localized modes, since their intensity profiles are identical [see Eq.~(\ref{modes2})] and accordingly the overlap with the input pulse is the same.
At the initial evolution stage, the radiation modes escape from the central region. Then, the superposition of two localized modes results in the total intensity oscillation
over $m$ with the period of $\Delta m=2$, see Figs.~\ref{fig:fig2}(b-d).
We extract the intensities of localized modes (see Sec.~3 of Supplemental Material \cite{SupplementalMaterial}) and plot with dots in Figs.~\ref{fig:fig2}(e-g). We observe a good agreement with theoretical predictions, except for a discrepancy at large $n$ in Fig.~\ref{fig:fig2}(e) due to the amplifier noise. We see that the localization becomes stronger for larger $\theta$, corresponding to a stronger cross-coupling between the loops and accordingly wider band-gaps as shown in Fig.~\ref{fig:fig2}(a).

\begin{figure}[htbp]
\centering
{\includegraphics[width= \linewidth]{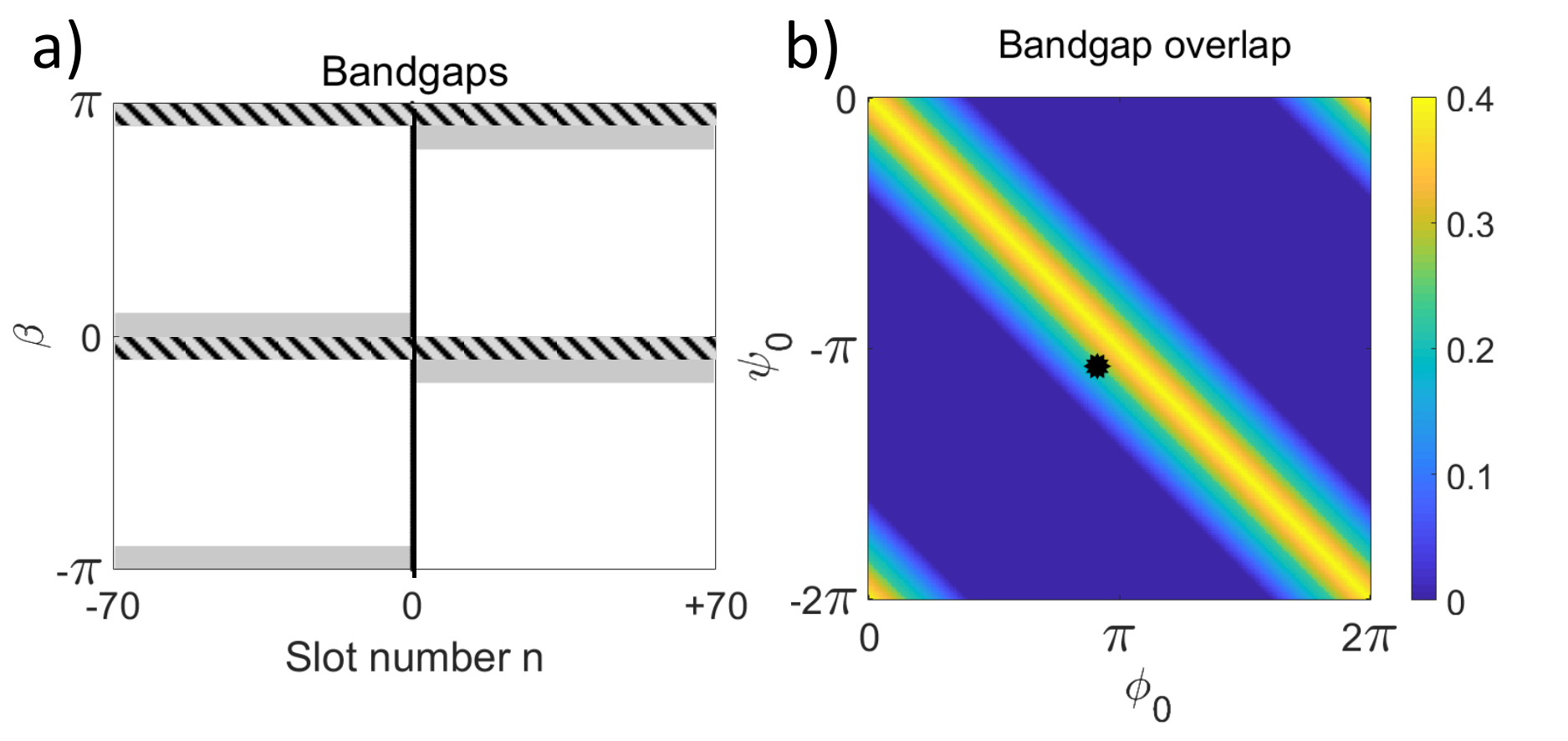}}
\caption{
a)~Bandgaps (shown with gray shading) and their overlap (shown with hatching) for regions with zero phases at $n<0$ and non-zero modulation at n$\geq$0 with $\phi_0=\psi_0+2\pi=0.9\pi$.
b)~Dependence of bandgap overlap on $\phi_0$ and $\psi_0$. The black dot corresponds to parameters in~(a).
For both plots, $\theta=0.1\pi$.
}
\label{fig:BandGapOverlapping}
\end{figure}

\begin{figure}[htbp]
\centering
{\includegraphics[width=\columnwidth]{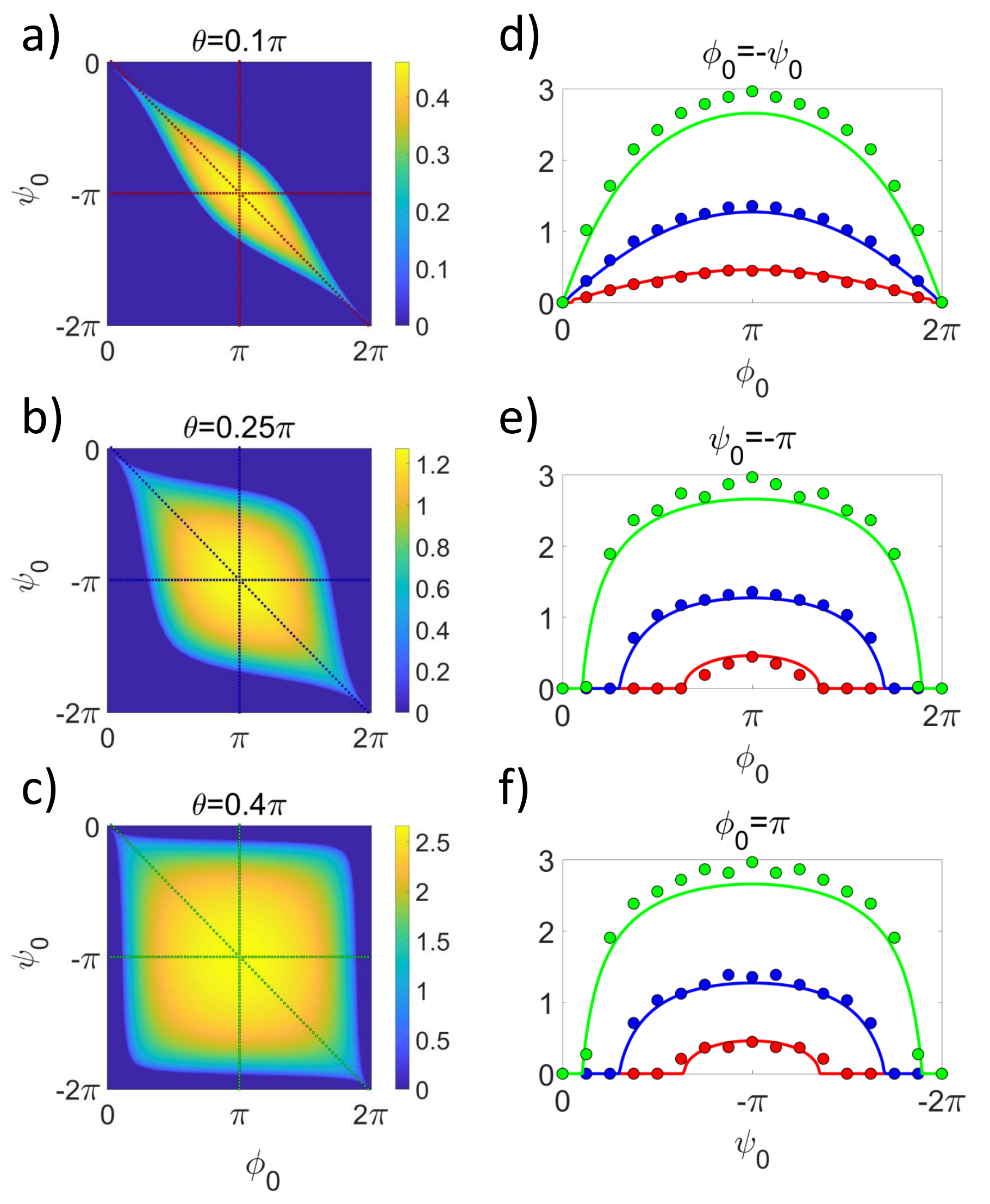}}
\caption{
Inverse localization width at intensity half-maximum.
(a-c)~Theoretical colormap vs. the phase modulations ($\phi_0$, $\psi_0$) for different coupling parameters $\theta$ as indicated by labels.
(d-f)~Experimental results (dots) in comparison with theoretical simulations (lines) for different cross-sections of panels (a-c): (d)~diagonal with $\phi_0=-\psi_0$, (e)~horizontal  with $\psi_0 = -\pi$, and (f)~vertical with $\phi_0 = \pi$. Red, blue, and green colors indicate $\theta=0.1\pi$, $0.25\pi$, and $0.4\pi$, respectively.
}
\label{fig:ExperimentComparison}
\end{figure}

We further investigate the robustness of the interface modes with respect to variations of the lattice parameters. The propagation constants of localized modes should be inside the band-gaps on both sides of the interface. For the out-of-phase modulation ($\phi_0 = - \psi_0$) considered above, the band-gaps exactly coincide. However in a more general case of $\phi_0 \ne - \psi_0$, the common band-gap becomes narrower as illustrated in Fig.~\ref{fig:BandGapOverlapping}(a). We present the dependence of the bandgap overlap width on the modulated phases in Fig.~\ref{fig:BandGapOverlapping}(b).
We numerically calculate the interface modes in the plane of phase modulations $(\phi_0,\psi_0)$, and plot
the inverse width at half-maximum of the localized mode intensity in Figs.~\ref{fig:ExperimentComparison}(a-c), for a set of coupling coefficients $\theta$. We observe that the localization occurs in a broad parameter range inside the bandgap overlap regions.
Such structural robustness of localization is distinctly different from the previously considered Aharonov-Bohm photonic caging at particular resonant conditions~\cite{Longhi:2014-5892:OL, Mukherjee:1805.03564:ARXIV}.

We perform systematic experimental measurements for sets of different phases $(\phi_0, \psi_0)$ and couplings $\theta$, and present the extracted inverse localization width for the observed localized modes with dots in Figs.~\ref{fig:ExperimentComparison}(d-f). We see an excellent agreement with theoretical predictions shown with lines. There results demonstrate that the modes always exist and are most strongly localized for a gauge field interface with $\phi_0=-\psi_0$, while localization persists in a broader parameter region which shape depends on the fiber loop coupling $\theta$.

In summary, we demonstrated
that a single discontinuity between uniform synthetic gauge fields can support localized modes.
This effect occurs in absence of band-gap shift or refractive index change, instead arising due to asymmetric phase accumulation similar to Aharonov-Bohm phenomenon.
We experimentally implemented a synthetic mesh lattice with real-time control over the vector gauge field, and observed strong localization which is structurally stable with respect to the control parameters.
We anticipate that such fundamental localization effect may appear in other types of physical systems with synthetic gauge fields, including waveguide lattices and coupled cavities for photons, cold atoms, and exciton-polaritons.

\begin{acknowledgments}
This work was supported by the Russian Science Foundation (16-12-10402). A.A.S. acknowledges support by the Australian Research Council (ARC) (DP160100619). I.D.V. acknowledges support of Ministry of Education and Science of the Russian Federation (3.7672.2017/8.9).
\end{acknowledgments}

\end{document}